\documentclass[aps,twocolumn,showpacs,prl,amsmath,amssymb,amsfonts,superscriptaddress]{revtex4}

\usepackage[varg]{txfonts}
\usepackage{color}
\usepackage{verbatim}
\usepackage{hyperref}
\usepackage{epsf}
\usepackage{graphics}
\usepackage{graphicx}
\usepackage{subfigure}
\usepackage[abs]{overpic}
\usepackage{natbib}
\usepackage{dcolumn}
\usepackage{bm}

\providecommand{\U}[1]{\protect\rule{.1in}{.1in}}

\begin{document}
\title{Interplay between spin-orbit coupling and Hubbard interaction in SrIrO3 and related Pbnm perovskites}
\author{M. Ahsan Zeb}
\email{maz24@cam.ac.uk}
\affiliation{Cavendish Laboratory, University of Cambridge,
Cambridge CB3 0HE, United Kingdom}
\author{Hae-Young Kee}
\email{hykee@physics.utoronto.ca}
\affiliation{Department of Physics, University of Toronto, Toronto, Ontario M5S 1A7 Canada }
\affiliation{Canadian Institute for Advanced Research, Toronto, Ontario, Canada}

\date{\today}
\begin{abstract}
There has been a rapidly growing interest on the interplay between
spin-orbit coupling (SOC) and Hubbard interaction $U$ in correlated materials.
A current consensus is that the stronger the SOC, the smaller is the critical interaction $U_c$
required for a spin-orbit Mott insulator,
because the atomic SOC splits a band into different total angular momentum bands narrowing the effective bandwidth.
It was further claimed that 
 at large enough SOC,
the stronger the SOC, the weaker the $U_c$ because in general the effective SOC is enhanced
with increasing electron-electron interaction strength.
Contrary to this expectation,
 we find that, in orthorhombic perovskite oxides (Pbnm), the stronger the SOC, the bigger the $U_c$.
This is originated from a line of Dirac node in $J_{eff}=1/2$ bands near the Fermi level inherited
from a combination of the lattice structure and a large SOC.
Due to this protected line of nodes, there are small hole and electron pockets in SrIrO$_3$, and  
such a small density of states makes Hubbard interaction less efficient in building a magnetic insulator.
The full phase diagram in $U$ vs. SOC is obtained, where non-magnetic semimetal, magnetic metal, and magnetic insulator 
are found.  Magnetic ordering patterns beyond $U_c$ are also presented. 
We further discuss implications of our finding in relation to other perovskites such as SrRhO$_3$ and SrRuO$_3$.
\end{abstract}
\maketitle

\section{Introduction}

Perovskite oxides with the chemical formula AMO$_3$ where A is a cation and M is a transition metal,
exhibit an exceptionally wide range of properties including 
anomalous Hall effect, 
colossal magnetoresistance, 
ferroelectricity, 
ferromagnetism, 
and superconductivity. 
Such an ample variety in a rather simple structure indicates that a detailed balance between
charge, spin, structure, and correlation is important in determining the ground state.

In particular, orthorhombic perovskite (point group symmetry, Pbnm) 
oxides are a large class of anisotropic oxides based on AMO$_3$ where  MO$_6$ octahedra 
are distorted from the symmetric cubic structure.
Among them, SrRuO$_3$, SrRhO$_3$ and SrIrO$_3$ (called perovskite ruthenates, rhodates, and iridates respectively),
display correlated metallic ground states. 
However, their magnetic properties differ hinting a crucial role of electron interaction.
SrRuO$_3$ is a ferromagnetic metal \cite{CaoPRB97,KatsPRB01,ChangPRL09} and SrRhO$_3$ a metal near a critical point \cite{YamauraPRB01,ShimuraJSSC92,SinghPRB03}, while SrIrO$_3$ is a semimetal 
with an extremely small number of charge carriers without any magnetic moment \cite{Takagi,LongoJSSC71,LiuMT05}. 
Given that Ir has 5d orbitals in the outer shell,
while Rh and Ru have 4d orbitals, Hubbard interaction is expected to be smaller in iridates \cite{RydenPRB70}.
Indeed it was found that the optical gap due to Hubbard interaction is about 0.5eV  in Sr$_2$IrO$_4$\cite{MoonPRB09}, 
a sister compound of SrIrO$_3$.
This leads to a naive conclusion that iridates should be better metal than rhodates or ruthenates, but the reality is the opposite.

What is missing in the above discussion is the SOC. 
Ir is heavier than Rh and the SOC strength is comparable to Hubbard interaction in iridates \cite{MoonPRB09}.
Since the atomic SOC is a local interaction, the electronic energy level splits into different total angular
momentum $J$ levels.
For example, starting from the atomic limit, five d-orbitals split into t$_{2g}$ and e$_g$
levels due to the octahedral crystal field, and t$_{2g}$ further splits into $J_{eff}=3/2$ and $J_{eff}=1/2$
via the SOC when the crystal field splitting is larger than the strength of SOC.
Once these bands form, a larger SOC leads to a smaller bandwidth of $J_{eff}=1/2$ separated from $J_{eff}=3/2$. 
Thus, the larger the SOC, the larger the ratio between Hubbard interaction($U$) and the bandwidth($W$), $U/W$
where $W$ is the bandwidth of $J_{eff}=1/2$.
While the absolute strength of $U$ is smaller in iridates,  its effect (given by the ratio $U/W$) is amplified.
This is indeed observed in a layered perovskite, Sr$_2$IrO$_4$, dubbed a spin-orbit Mott insulator \cite{BJKimPRL08,BJKimScience09,JinPRB09,WatanabePRL10,MartinsPRL11,AritaPRL12,FujiyamaPRL12}.
To explain the metallicity of SrIrO$_3$ compared to insulating Sr$_2$IrO$_4$, it was further suggested that SrIrO$_3$
has a larger bandwidth comparing to quasi-two dimensional Sr$_2$IrO$_4$ \cite{MoonPRL08,LiuPRL08}.
A growing consensus is that the larger the SOC, the smaller the critical interaction strength $U_c$ that is required for
the phase transition from metal to Mott insulator \cite{WatanabePRL10,PesinNP10}

However, once the SOC splits the t$_{2g}$ bands into different $J_{eff}$ bands,
its effect on the bandwidth of $J_{eff}=1/2$ is minimal,
and the interplay between the SOC and the electron-electron interaction is intriguing.
It was claimed that in general the effective spin-orbit coupling is enhanced
with increasing strength of the electron-electron interaction leading to the same conclusion that
a larger SOC leads to a smaller $U_c$.\cite{PesinNP10,LiuPRL08}

%

\begin{figure}[t]
\includegraphics[width=0.6\columnwidth]{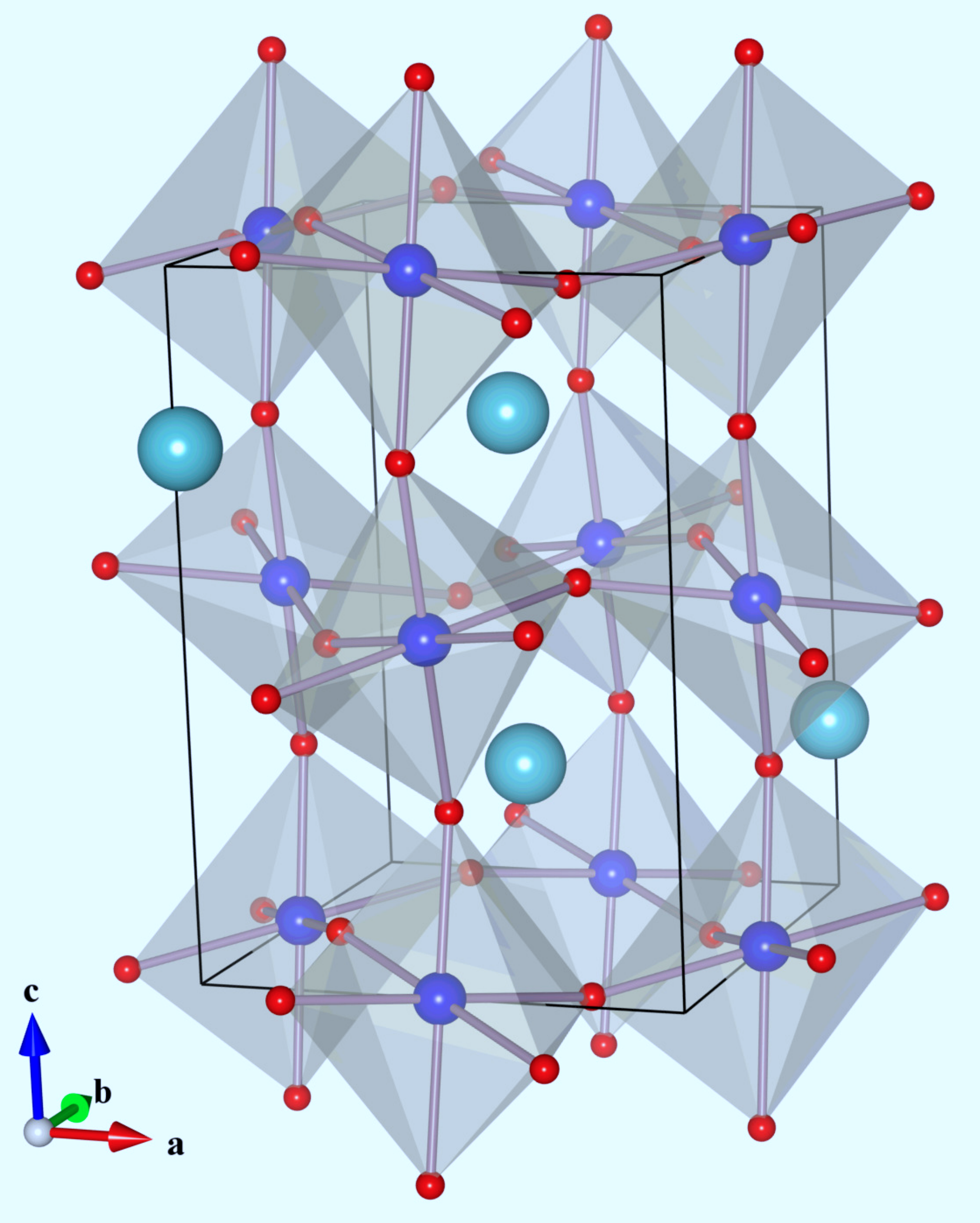}
\caption{\label{xl}{
Crystal structure of orthorhombic perovskite SrIrO$_{3}$.
Sr, Ir and O atoms are shown in aqua, blue and red. 
The octahedra shown are rotated about  the $z-$axis and tilted
about 
 $[110]$-axis making the unit cell four time
 bigger than that of the cubic perovskite structure.
 }}
\end{figure}

In this paper, we show a counter example where the common wisdom does not apply. 
We study the interplay between SOC and Hubbard interaction in orthorhombic perovskite oxides (Pbnm).
It is found that the bigger the SOC, the larger the $U_c$ 
in orthorhombic perovskites inherited to
 the lattice structure.
When SOC is moderate 
(close to the true SOC in SrIrO$_3$),
the band dispersion exhibit a line of Dirac node protected by the symmetry of the lattice.
We propose that semimetallicity in SrIrO$_3$ compared to insulating Sr$_2$IrO$_4$ is due to
such a small density of states, which in turn requires a larger $U_c$ for the transition to a Mott insulator. 
Hubbard interaction in iridates is smaller than this $U_c$, and thus SrIrO$_3$ remains metallic with small Fermi pockets .
Beyond $U_c$,  non-collinear and non-coplanar magnetic structures appear, and the overall phase diagram contains 
ferromagnetic metal, non-magnetic semimetal and magnetic insulator.
Below we will show the band structures computed for SrIrO$_3$, 
where we use Hubbard $U$ and SOC strength
$\alpha$ as tuning parameters to understand different phases realised in other orthorhombic perovskite oxides
such as SrRuO$_3$ and SrRhO$_3$.  Our findings 
suggest that the SOC together with Hubbard interaction $U$ plays an important role in realising different ground states in 
SrRuO$_{3}$ \cite{ChakoumakosPB98,HamlinPRB07,AlexanderPRB05,KimPRB03,AhnPRL99,KeAPL04},  SrRhO$_{3}$ \cite{YamauraPRB01}, and SrIrO$_3$ \cite{CaoPRB07,Laguna-MarcoPRL10,JangJPCM10}.

The paper is organised as follows. 
In the following section, the details about the crystal structure is presented.
In Sec. 3, computational method is explained, and the band structures and phase diagram
in $U$ vs. SOC  are presented in Sec. 4. Magnetic metal and insulator appear at small and large $U$ respectively,
and their magnetic ordering patterns depend on the SOC which will be shown in Sec. 5. 
A brief summary and implications of our findings are listed in the final section.

\section{Crystal structure}

Fig.\,\ref{xl} shows the crystal structure of the orthorhombic perovskite SrIrO$_{3}$ with 
Sr, Ir and O atoms as aqua, blue and red balls.
As can be seen in Fig.\,\ref{xl}, the octahedra enclosing the Ir atoms
are rotated about the $z-$axis and tilted about 
 $[110]$-axis.
Due to these rotations and tilts, there are four formula units of SrIrO$_{3}$ in a unit cell and the octahedra also get distorted. 
For any two connected octahedra,
 the rotations are in the same (opposite) direction if the two enclosed Ir atoms lie in different (the same) layers, 
whereas the tilts are opposite for all nearest neighbour octahedra. 
 
The experimental lattice parameters of this Pbnm phase of SrIrO$_{3}$ are
 $a=10.5136$ a.u., $b=10.5688$ a.u. and $c=14.9$ a.u.,
and an asymmetrical unit is: a Sr at $(0.5085,0.4901,0.25)$, an Ir at $(0.5,0,0)$ and
two O at $(0.506,0.073,0.25)$ and $(0.292,0.714,0.044)$ \cite{ZhaoJAP08}.

This structure is primitive orthorhombic for which the symmetry elements include
two $\mathit{b}$ glide planes perpendicular to $x$-axis at  $x/a=1/4$ \& $3/4$,
two $\mathit{n}$ glide planes perpendicular to $y$-axis at  $y/b=1/4$ \& $3/4$ and 
two mirror planes perpendicular to $z$-axis at  $z/c=1/4$ \& $3/4$.
Here, a $\mathit{b}$ ($\mathit{n}$) glide plane means that a reflection across the plane followed by a translation
 of $\mathbf{a}/2$ ($[\mathbf{a+c}]/2$, i.e, along the diagonal) transforms the structure to self coincidence.
Furthermore, there are four $2_{1}$ screw axes parallel to each of the three primitive lattice vectors $\mathbf{a}$,$\mathbf{b}$ and $\mathbf{c}$.
The $2_{1}$ screw axes parallel to $\mathbf{a}$ or $x$-axis are at $(y/b,z/c)=(1/4,0),(1/4,1/2),(3/4,0)$ \& $(3/4,1/2)$; those parallel to $\mathbf{b}$ or $y$-axis are at $(x/a,z/c)=(1/4,1/4),(1/4,3/4),(3/4,1/4)$ \& $(3/4,3/4)$; and those parallel to $\mathbf{c}$ or $z$-axis are at $(x/a,y/b)=(0,0),(0,1/2),(1/2,0)$ \& $(1/2,1/2)$. 

There are eight inversion centres at $x/a,y/b,z/c \in \{0,1/2\}$.
The four Ir atoms in the unit cell sit at four of these. 
This also means that all the octahedra in Fig.\,\ref{xl} are inversion symmetric. 
While this is obvious in case of a cubic perovskite structure which forms regular octahedra around the Ir atoms, it is not so in this case where the octahedra are distorted.
Two of the $2_{1}$ screw axes parallel to $\mathbf{c}$ passes through the Ir atoms. These screw axes and the four inversion centres at Ir locations are necessary for the existence of the mirror planes at $z/c=1/4$ \& $3/4$, which connect the octahedra in two different layers through the reflection symmetry. 
It was found in Ref. \cite{CarterPRB12} that  breaking this mirror plane symmetry is a way to generate a strong topological insulator.


\begin{figure}[!ht]
\subfigure[]{
\begin{overpic}[width=0.92\columnwidth]{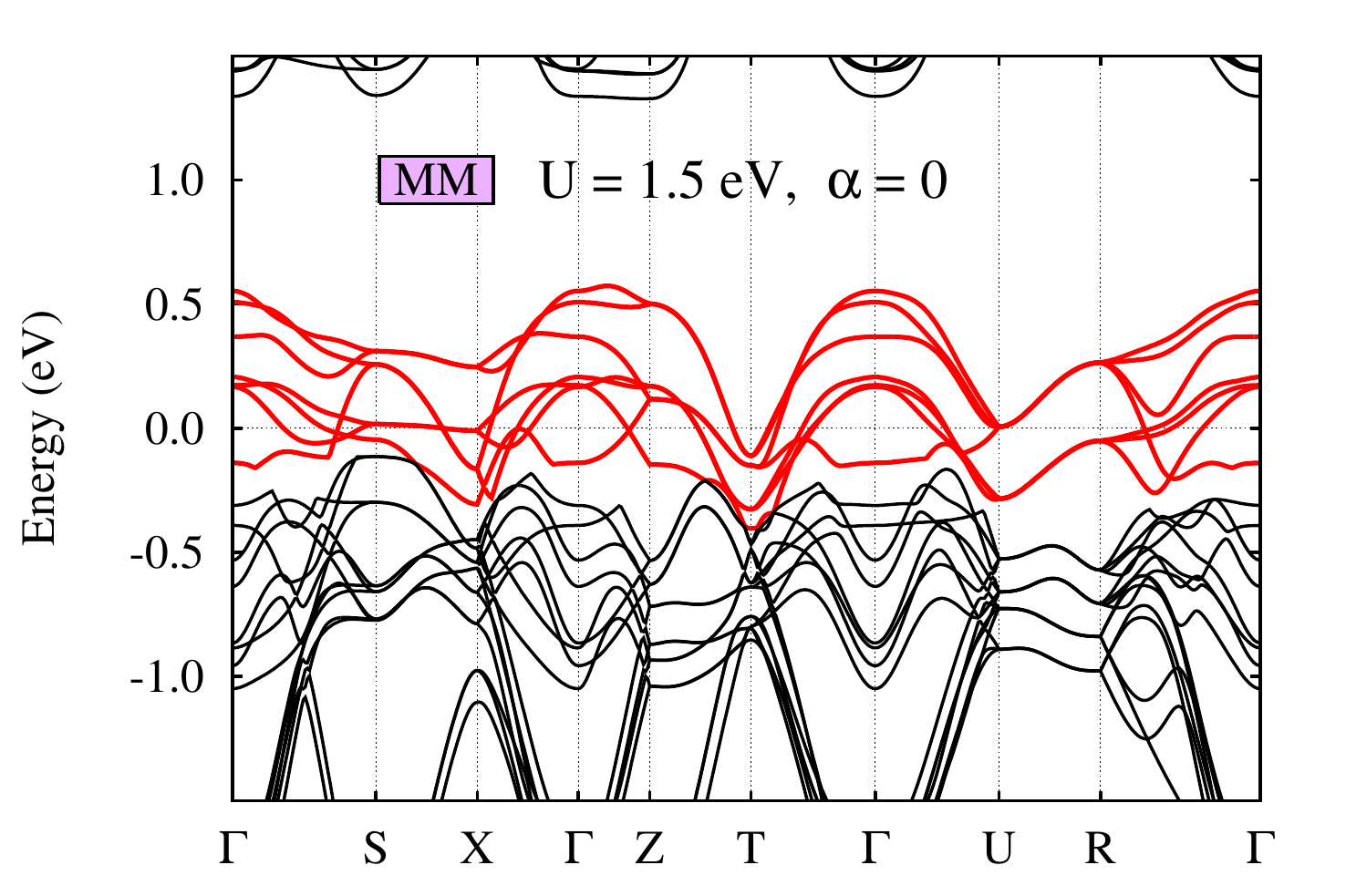}
\end{overpic}
}
\subfigure[]{
\begin{overpic}[width=0.92\columnwidth]{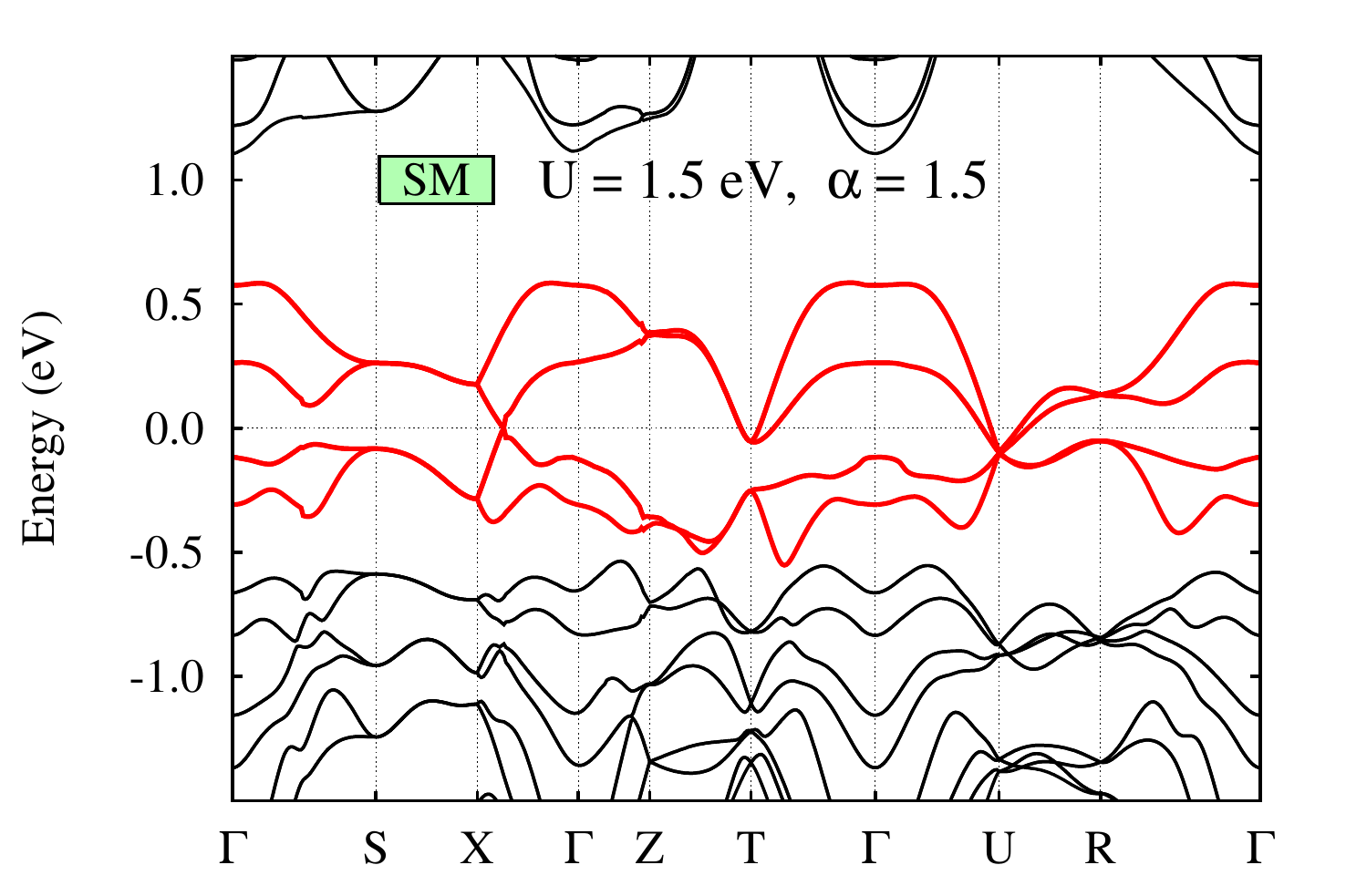}
\end{overpic}
}
\subfigure[]{
\begin{overpic}[width=0.92\columnwidth]{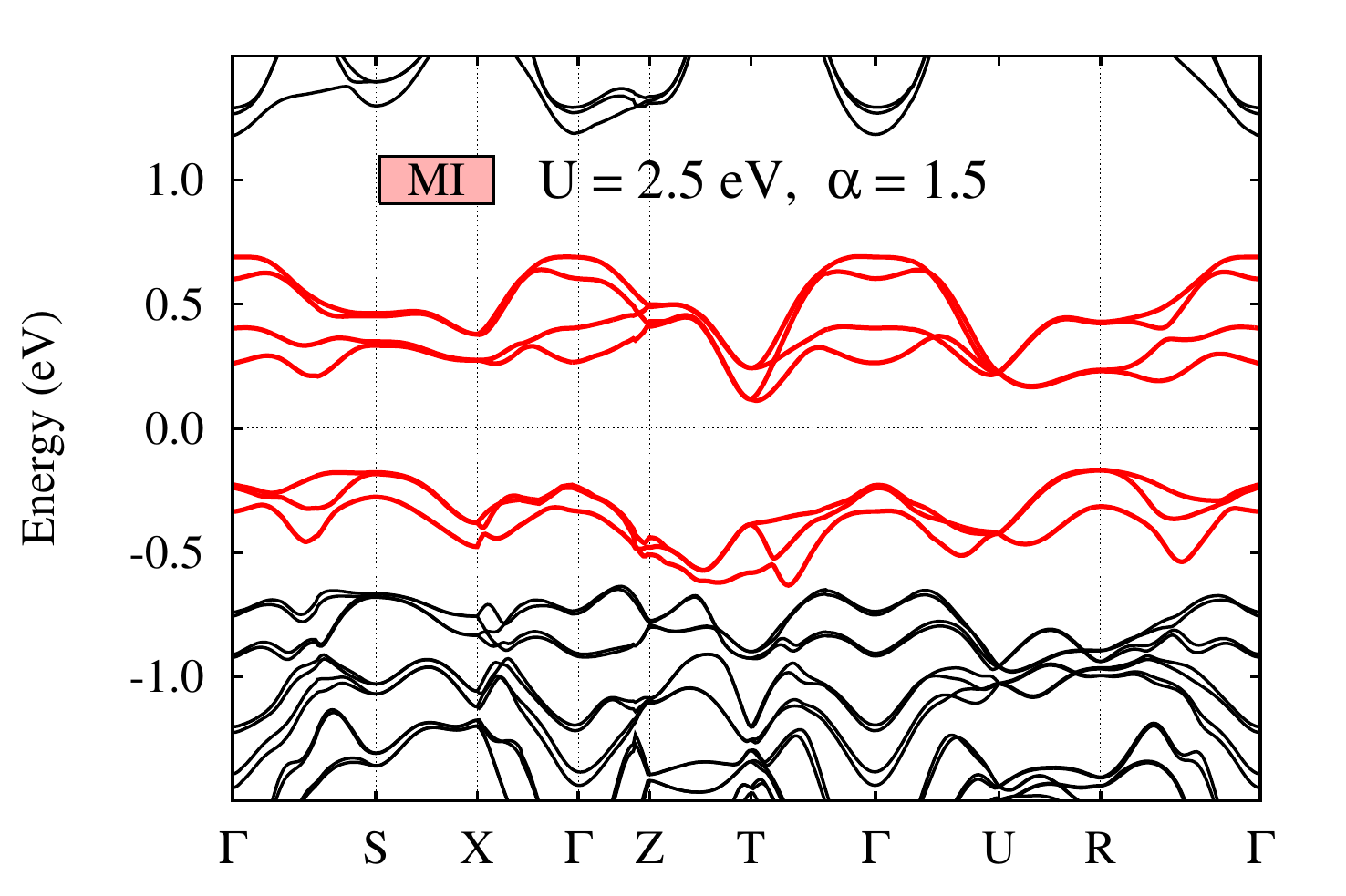}
\end{overpic}
}
\caption{\label{rep-bands}{
Some representative band structure diagrams of orthorhombic perovskite oxides
for (a) magnetic metal (MM) at $U=1.5$ eV and $\alpha=0$,
 (b) semimetal (SM) at $U=1.5$ eV and $\alpha=1.5$, and 
 (c) magnetic insulator at $U=2.5$ eV and $\alpha=1.5$.
 The bands near the Fermi energy are denoted by red color,
  and remains knotted near U in SM phase.
 }}
\end{figure}

\section{First principle calculations}
We performed density functional theory (DFT) \cite{HohenbergPR64,KohnPR65} calculations 
including Hubbard $U$ and SOC 
using the full-potential linearised augmented-plane-wave (FP-LAPW) method as 
implemented in the elk code \cite{elk}.
The local density approximation (LDA) for the exchange and correlation energy functional
 in the Ceperley-Alder \cite{CeperleyPRL80} form parametrized by Perdew and Zunger \cite{PerdewPRB81} was employed. 
We 
used the ``around mean field" (AMF) scheme \cite{CzyzykPRB94} for the double-counting-correction.
That is, to correct our DFT+U calculations for the Coulomb repulsion already present
  in the DFT Hamiltonian.
  We treated up to $3d$ of Sr, up to $5s$ of Ir excluding $4f$, and $1s$ of O
with the radial Dirac equation, while the scaler relativistic approximation is used to include the SOC for the higher states in the second variational step \cite{KoellingJPCSSP77}. 

To confirm that our main results are robust to the choice of double counting correction,
we have also computed the band structures using the ``fully localized limit" (FLL) correction
\cite{Anisimov93,Liechtenstein95,YlvisakerPRB09} near the phase boundary. We found that at large SOC,  $U_c$ is essentially the same. 
However, for small SOC, $U_c$ is shifted towards a lower value than that found with AMF correction 
in such a way that our main conclusion (the larger the SOC, the larger the $U_c$) does not alter. 
The phase boundaries obtained by these two different corrections are denoted by different colours in 
the phase diagram shown in Fig.\,\ref{phase}, and will be discussed below.

To obtain the phase diagram of SrIrO$_{3}$ in the $U$-SOC phase space, we 
tune the SOC term for the $5d$ orbitals of Ir atoms.
Since the strength of the
 SOC increases sharply with the atomic number $Z$ ( as $Z^{4}$)
 , it is much stronger for Ir ($Z=77$) as compared to Sr ($Z=38$) or O ($Z=8$).  This means that Ir contributes almost exclusively to the SOC energy in SrIrO$_{3}$. 
This allows us to safely tune the SOC for all valence states, because its effect on Sr and O atoms does not count much.
A scaling factor $\alpha$ in the SOC term of the Hamiltonian is introduced in the second variational step \cite{KoellingJPCSSP77}. 
This way, we can enhance the effect of SOC by taking $\alpha>1$ or reduce it by taking $\alpha<1$.
For instance, $\alpha = 0$ would mean no SOC at all, while $\alpha =1$ is the atomic SOC in Ir atoms. 
A small magnetic field is used to set the quantisation direction for the angular momentum.
 This field reduces exponentially to zero during the self consistency iterations so it has no other effects. 

 In the FP-LAPW method, the real space is divided into spheres around the atoms (muffin-tins) and interstitials elsewhere. 
In the present calculations, the muffin-tin radii 
 $1.86$ a.u. , $2.08$  a.u.  and $1.51$  a.u. are used
 for Strontium (Sr), Iridium (Ir) and Oxygen (O), respectively.
The basis set consists of APW functions with angular momentum $l$ up to $8$ and plane waves with cut-off energy equal to $231.3$ eV. 
The number of empty states in the basis set in the second variational step was $10$. 
The Brillouin Zone integrations were performed using a $3\times3\times3$ grid, 
which is equivalent to using $10$ points in the irreducible part of the Brillouin Zone. 
This works well, given that the primitive unit cell of orthorhombic perovskite SrIrO$_{3}$
 is almost four times bigger than that of the cubic structure with only one formula unit.
We checked the k-grid convergence in the metallic phase using $8 \times 8 \times 8$ grid.
We only used $U$ for $5d$ orbitals of Iridium.

\section{Band structures and phase diagram}

The octahedral crystal field splits the bands derived from the $d$-orbitals of transition metal atoms into high energy $e_{g}$ 
and low energy $t_{2g}$ groups.
Due to the distortion of octahedra, there are twelve $t_{2g}$ and eight $e_{g}$ bands 
(each band is doubly degenerate due to time reversal symmetry).
Fig.\,ref{rep-bands} shows band structures for various values of $U$ and SOC denoted in the inset.
The crystal field gap between $e_{g}$ and $t_{2g}$ is evident for all cases, 
and only bottom of $e_{g}$ bands are shown in the plots. 

When $\alpha =0$ that corresponds to the absence of SOC, a ferromagnetic order is present, and  
 $t_{2g}$ bands are all mixed as shown in panel (a). In contrast, 
when  $\alpha = 1.5$, panel (b) and (c), the $t_{2g}$ bands form two groups, 
the higher four half filled bands originate from the $J_{eff}=1/2$ denoted by red colour (
the lower two bands near $\Gamma$ point are mainly $J_{eff}=3/2$ though), and the lower 
eight completely filled bands from the $J_{eff}=3/2$.
Increasing the SOC increases the splitting between the $J_{eff}=1/2$ and $J_{eff}=3/2$ bands.

\begin{figure}[!ht]
\begin{center}
\includegraphics[width=0.5\textwidth]{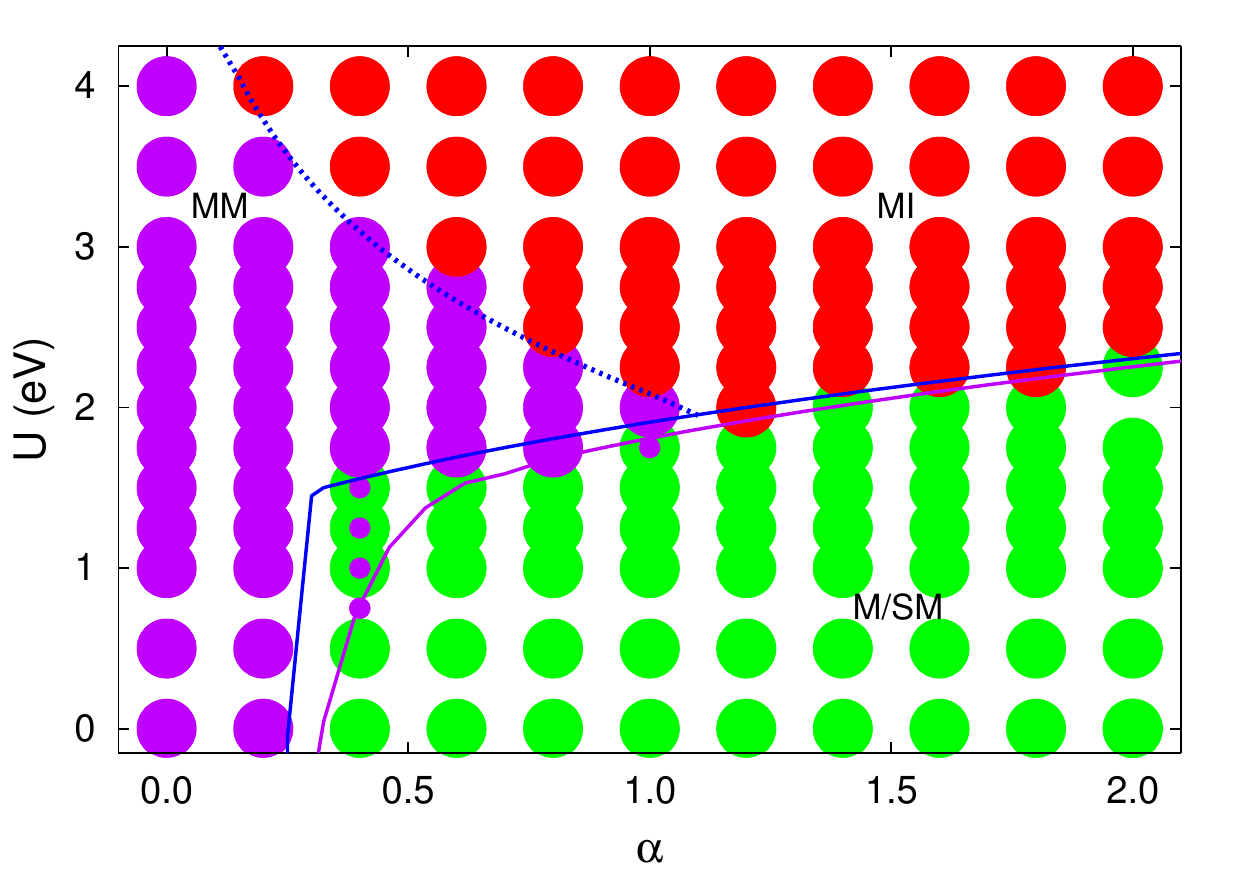}
\caption{\label{phase}{
The phase diagram of orthorhombic perovskite oxides in the $U$-SOC plane.
Three phases for $U$ up to $4$ eV and $\alpha=0-2$ are Magnetic Metal (MM),
non-magnetic Metal or Semimetal(M/SM), and Magnetic Insulator (MI).
The colour circles show the points for which calculations have been performed
and  magenta, green and red denote MM, M/SM, and MI, respectively. 
Small magenta circles are FLL results showing MM phase where AFM gives M/SM phase.
The solid line separates two phases connected via a first order phase transition
(where blue line is obtained by AMF while magenta line is by FLL),
whereas the dotted line is the phase boundary for a second order phase transition.
}}
\end{center}
\end{figure}

For smaller $U$ of panel (b), the phase is nonmagnetic semimetal, where four $J_{eff}=1/2$
bands are near the Fermi level forming small pockets of Fermi surface.
While in non-magnetic semimetal (SM) and magnetic metal (MM) phases, panel (a) and (b), 
there is a finite density of states at Fermi energy, 
the band topologies are very different in these two phases.
In the non-magnetic semimetallic (SM) phase, the bands at the fermi energy cross near U point resulting in a line node, and 
the magnetisation is zero everywhere in this phase.
In the MM phase, as well as in the magnetic insulator (MI) phase shown in panel (c),
there is no such band crossing.
In both these phases, Ir atoms have finite magnetic moments with a long range order. 
Increasing $U$, 
keeping the same strength of SOC, leads to a metal-insulator
transition at a critical $U_c$, where the insulating state such as panel (c) has an interesting magnetic ordering pattern. Since the time reversal symmetry is broken due to the magnetic ordering, there are eight $J_{eff}=1/2$ bands
in this phase as displayed in panel (c). A further discussion about the magnetic ordering pattern will
be presented below.

These three electronic phases shown in Fig.\,\ref{rep-bands} are found in the $U$-SOC phase diagram;
$(i)$ M/SM, $(ii)$ MM, and $(iii)$ MI.
The overall phase diagram in $U$ vs. SOC is presented in Fig.\,\ref{phase},
where M/SM, MM, and MI are shown in green, magenta, and red.
M/SM is connected to MM and MI via a first order phase transition
whereby the magnetisation jumps from zero to a finite value along with a sudden change in the band structure topology. 
On the other hand, MM and MI transform into one another continuously with opening or closing up of a band gap.


Let us discuss the phase diagram by checking along different cuts. First vertical cuts, i.e., changing $U$ for a given $\alpha$.
When $\alpha=0$, the system remains a pure ferromagnetic metal at all $U$.
This results
from a large density of states at the Fermi level leading to a Stoner ferromagnet.
Tuning SOC to finite but still small values (for $\alpha < 0.3$), $U$ interaction does not make any  difference,
and system stays in the magnetic metal phase even for very high values of $U$ (for $U$ close to 5eV, it becomes ferromagnetic insulator, which is not shown here).
However, as the SOC does not favour a pure ferromagnetic ordering, it turns the magnetic ordering pattern to a slightly
non-coplanar order with a large ferromagnetic component. 
In contrast, for $ \alpha > 0.3 $, increasing $U$ induces a first order phase transition from 
non-magnetic semimetal to magnetic phases.  Whether the magnetic phase is metal or insulator depends on the strengths of both $U$ and $\alpha$.
The phase boundary separating the non-magnetic semimetal phase from the two other phases, MM and MI, is shown 
as a solid line in Fig.\,\ref{phase}.
For $\alpha > 1.1$, increasing $U$ transforms M/SM directly to MI, while for $0.3 < \alpha < 1.1$, increasing $U$ changes the phase 
from non-magnetic metal, to magnetic metal followed by magnetic insulator.

Let us explore the phase diagram using horizontal cuts -- changing $\alpha$ for a fixed $U$.
For small $U$, increasing $\alpha$ leads to a first order phase transition from magnetic metal to non-magnetic
metal/semimetal phase. The critical value of $\alpha$, $\alpha_{c}$, 
 at which this transition takes place stays between $0.2-0.4$ for $0\leq U<1.5$ eV.
 This is rather expected, as SOC disfavours spin density wave ordering
within a weak coupling theory. 
Fig.\,\ref{phase} shows results for $\alpha$ up to $2$. 
As can be seen, there is no further phase transitions by increasing $\alpha$. 
We checked this for $\alpha$ up to $5$.

For $U\geq 1.5$ eV, $\alpha_{c}$ increases sharply with $U$
 with an increasing separation between the bands at the Fermi level in magnetic metal phase.
It is also interesting to note that for $2\leq U\leq 2.35$ eV, the system undergoes
a change in phase
by increasing $\alpha$
 from magnetic metal to magnetic insulator, 
and then into non-magnetic semimetal,
 i.e, a re-entrance of metallicity (metal-insulator-metal by
changing SOC for a given $U$).
 For $U\geq 2.35$ eV, increasing $\alpha$ transforms magnetic metal smoothly to magnetic insulating phase
 with opening up of a band gap. 
 The higher the value of $U$, the lower is the value of $\alpha$ for this transition.
 The phase boundary between these two phases is shown as a dotted line in Fig.\,\ref{phase}.

\begin{figure}[t]
\begin{center}
\includegraphics[width=0.3\textwidth]{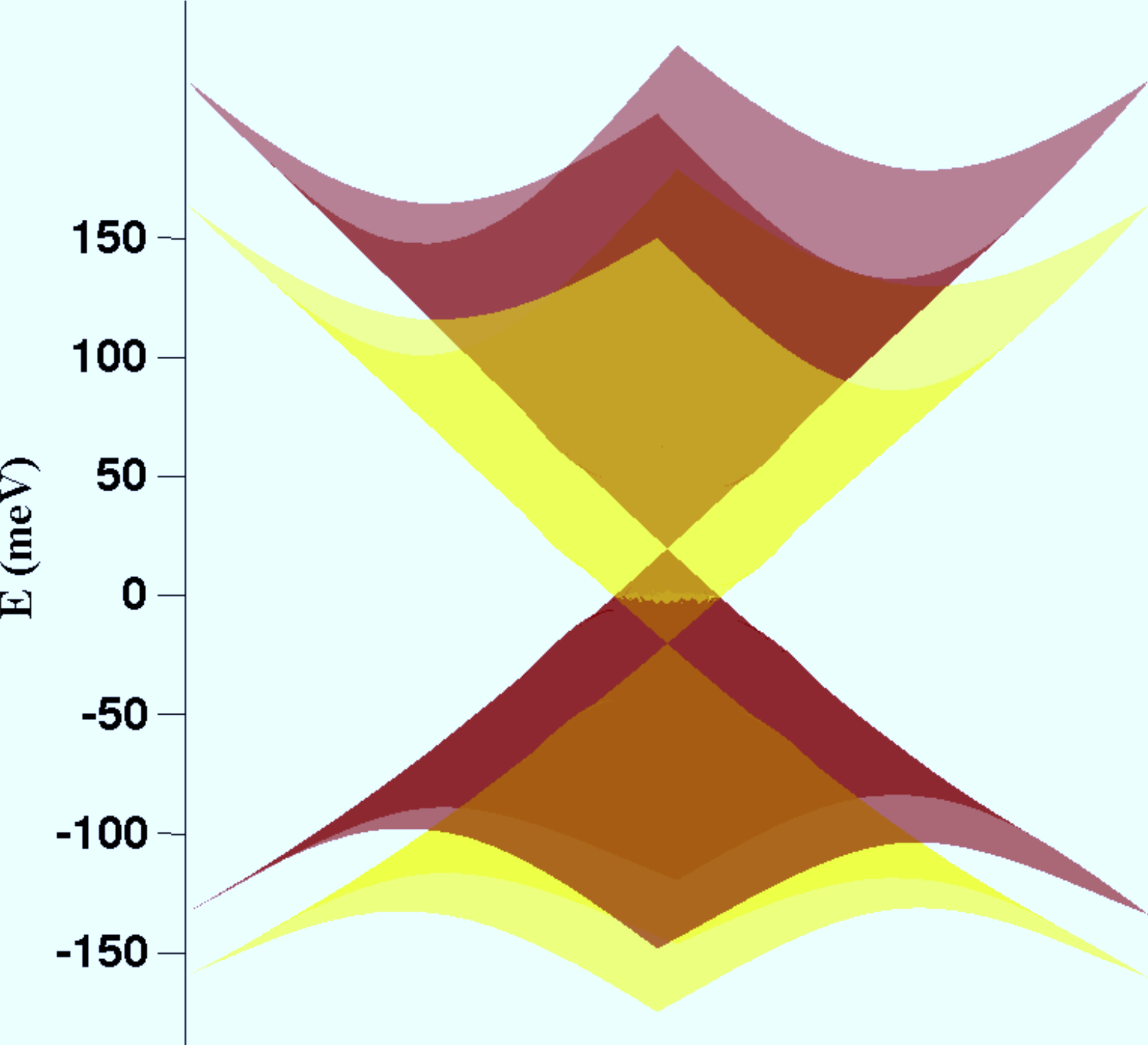}
\caption{\label{band}{
The $J_{eff}=1/2$ bands of SrIrO$_{3}$  close to U in the XURS-plane computed with LDA (without +U)
at $\alpha = 1$ (atomic SOC). 
These bands form two pairs of touching cones shown in yellow and brown.
The two pairs are interpenetrated into each other forming a circular nodal line at the fermi energy.
When the time reversal symmetry is broken (as the magnetic ordering occurs), these band crossings 
disappear and a band gap form.
}}
\end{center}
\end{figure}

As can be seen in Fig.\,\ref{phase},
the dotted line have the expected phase boundary curvature in the $U$-SOC plane. That is, the critical value of $U$ 
for MM to MI transition decreases with increasing $\alpha$ since the phase space of magnetic insulator should be wider
as $U$ increases. 
However, the most unexpected result presented in Fig.\,\ref{phase} is  the opposite phase boundary curvature
for transition to magnetic phases from non-magnetic metallic phase, the solid line.
It shows that a stronger SOC requires a stronger electron-electron interaction
 to transform non-magnetic metal to magnetic insulator.
The origin of this unexpected behaviour is likely to be the special band topology in the semimetal phase as described below.


 Fig.\,\ref{band} shows 
the band structure near U in the XURS-plane.
The four $J_{eff}=1/2$ bands form two interpenetrated pairs of cones, each pair consisting of a lower and a higher band (a yellow and a brown).
One pair touches below the fermi level while the other above it, forming two Dirac-like points and a circular line node in the XURS-plane at the fermi energy. 
Due to this node, there is an extremely small density of states near the Fermi level,
which 
in turn requires
  a high Hubbard $U$ to splits
these cones resulting in a magnetic insulator.
We propose that this is the main mechanism whereby SrIrO$_3$ is a semimetal with a small carrier density, different from its sister compounds Sr$_2$IrO$_4$ \cite{BJKimPRL08,CrawfordPRB94,ShimuraPRB95,KimPRL12} and Sr$_3$Ir$_2$O$_7$ \cite{CaoPRB98,CaoPRB02,NagaiJPCM07}.

\begin{figure}[t]
\begin{center}
\includegraphics[width=0.5\textwidth]{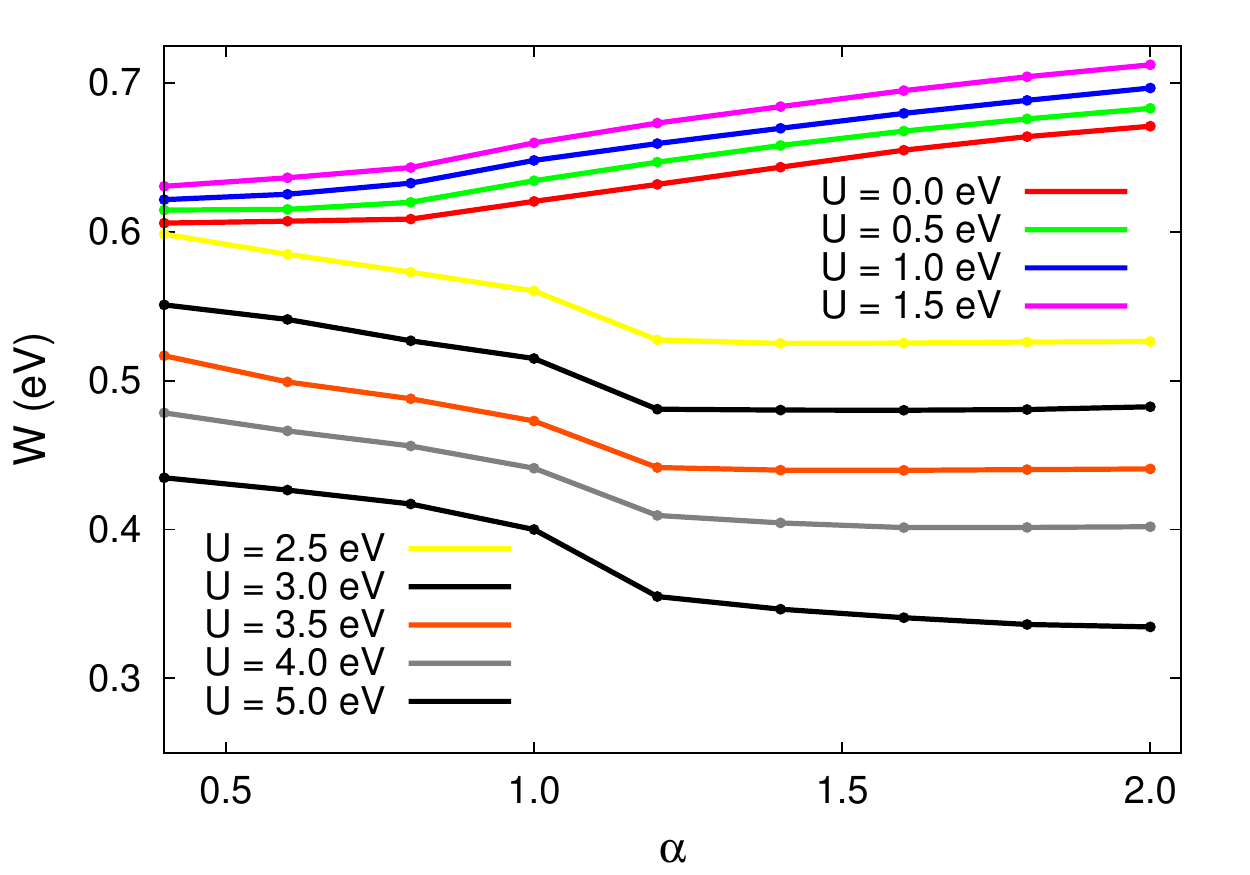}
\caption{\label{wsoc}{
The bandwidth W of top two bands at fermi level as a function of $\alpha$ for 
various values of $U$. W decreases with $\alpha$ and $U$  in MM/MI phase ($U\geq2.5$ eV)
 as expected but increases in SM phase ($U\leq1.5$ eV).
 }}
\end{center}
\end{figure}

We also checked the bandwidth of the upper two bands of $J_{eff}=1/2$ at the Fermi level
when SOC is large enough to separate $J_{eff}=1/2$ top two bands from the rest (except at the nodal points).
W is plotted in Fig.\,\ref{wsoc} against $\alpha$ for various values of $U$ for $\alpha \ge 0.5$,
since $\alpha < 0.5$, $J_{eff} =1/2$ is not well defined.
Contrary to the expectation \cite{MoonPRL08},
the bandwidth increases with $\alpha$ and $U$ in the non-magnetic semi-metallic phase
 most likely due to a steeper slope of Dirac node that confirms our conclusion above.
Whereas in the magnetic phases (the plots for $U=2.5-5$ eV), W decreases with $\alpha$ and $U$ as expected.
This makes us believe that the transition from the non-magnetic to magnetic phases
in SrIrO$_{3}$ is controlled by the electronic state of the non-magnetic semimetal
where the bandwidth is not relevant.  This is further supported by the fact that the semimetal phase has a
special band topology as described below.


J. M. Carter {\it et al} showed in Ref. \cite{CarterPRB12} using a tight binding model that
this line node is proteced by the lattice symmetry. In other words, any term
that opens up a gap near U-point should break either time reversal, inversion, or Pbnm lattice symmetry.
For example, it was shown that when the mirror symmetry between the two layers of
IrO$_{2}$ in SrIrO$_{3}$ is broken with a staggered potential,
this line node changes to a point node.
When the strength of this staggered potential is increased beyond a critical value
that takes the node to the R point, a change in the topology of the bands occurs  owing to the inversion of the $J_{eff}=1/2$ bands at R.
The system is turned into a strong topological insulator when this happens.
Further increase of this staggered potential leads to inversion of the bands at Z point, changing the band topology back to trivial,
making the system a band insulator.

\section{Magnetic ordering patterns}

As discussed above, when  $\alpha=0$, there is a pure ferromagnetic (FM) order in the MM phase at all values of $U$. 
This happens down to $U=0$,  because the electron-electron repulsion is not completely absent even at $U=0$, partly due to its imperfect removal in the $5d$ orbitals of Ir and partly due to the presence of many other occupied states in the system.
The magnitude of the magnetic moment of Ir depends on $U$.
It increases with $U$ from $0.38 \mu_{B}$ at $U=0$ to $0.95 \mu_{B}$ at $U=4$ eV,
where it is almost saturated --- a rather expected behaviour.
A small contribution to the magnetisation also comes from O when $\alpha\sim0$.

\begin{figure}[t]
\begin{center}
\includegraphics[width=0.5\textwidth]{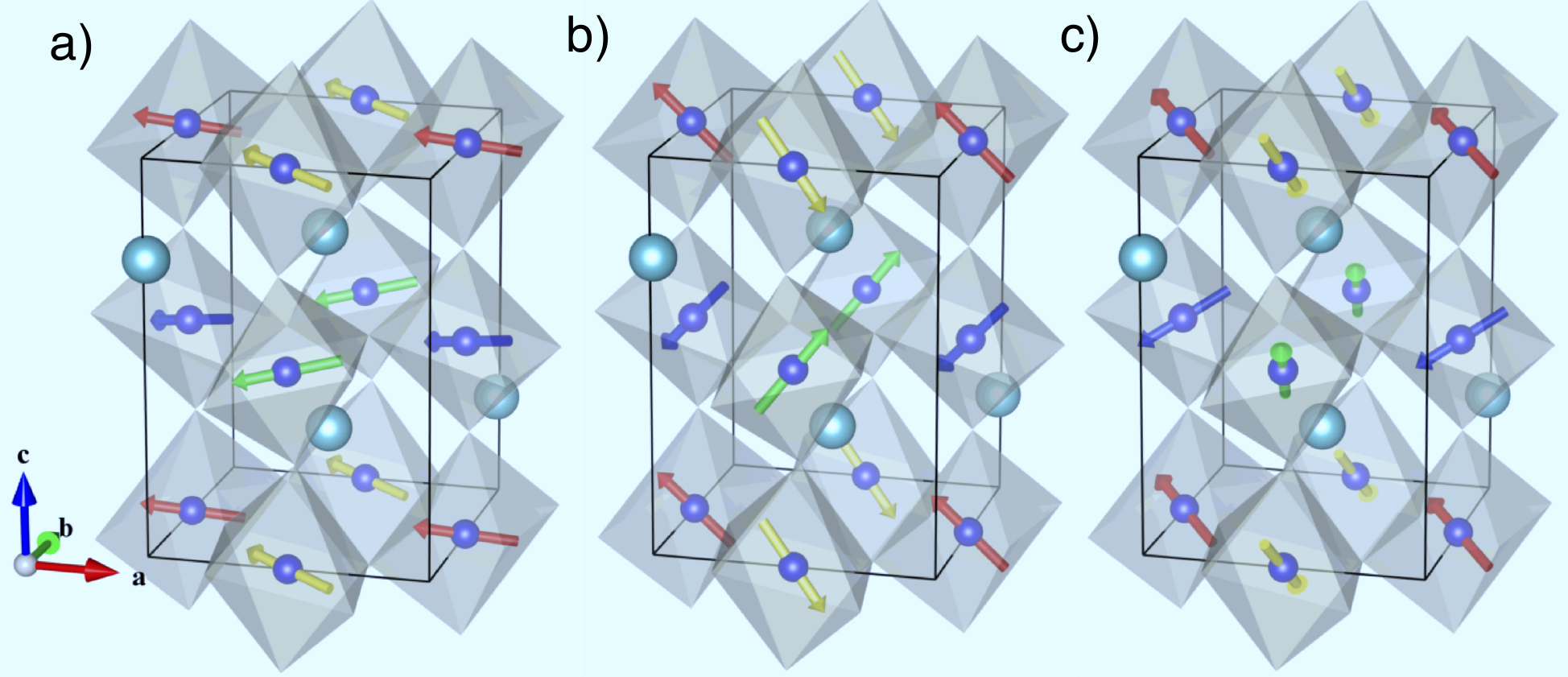}
\caption{\label{ma}{
The magnetic structure of orthorhombic perovskite oxides at 
$U=2$ eV and $\alpha=0.2$ (a),
$U=4$ eV and $\alpha=0.2$ (b),
$U=4$ eV and $\alpha=1$ (c).
In (a), the system has a canted antiferromagnetic order with a 
large ferromagnetic component.
In (b), the system has a canted antiferromagnetic order with negligible ferromagnetic component.
The average magnetisation per unit cell is also zero in (c), but the moments are aligned in a very different way.
 }}
\end{center}
\end{figure}

As we move away from $\alpha=0$, system develops a 
canted antiferromagnetic (CAFM) order. 
At smaller $U$, transition from FM to CAFM is more gradual leaving a net ferromagnetic component. 
This behaviour persists up to $\alpha \sim 1$. For higher $\alpha$ a very small ferromagnetic component develops in the 
magnetic insulator phase at higher $U$. 
This is expected since an antiferromagnetic order in the insulating phase 
lowers the energy via virtual hopping of electrons to the nearest neighbour with the oppositely aligned spin.
A small ferromagnetic component is then due to an effective Dzyalonshinsky-Moriya interaction as found in Sr$_2$IrO$_4$ \cite{BJKimScience09,BJKimPRL08,JackeliPRL2009}. 
Fig.\,\ref{ma} shows the magnetic structure
at $\alpha=0.2$ as $U$ is changed from $2$ to $4$ eV,
and at $U=4$ eV as $\alpha$ is changed from $0.2$ to $1$.
The quantisation axis is set along the $x$-axis.
 
For any two nearest neighbour Ir atoms in different layers along $\mathbf{c}$
(i.e, those with yellow and green or red and blue arrows in Fig.\,\ref{ma})
the components of moments along the $y$ and the $z$-axis
are always cancelled out.
The size and orientation/direction of individual moments depend on 
the values of $\alpha$ and $U$ as does their sum or the total moment per unit cell.
As can be seen in Fig.\,\ref{ma}(a),
at $U=2$ eV and $\alpha=0.2$,
the moments are almost co-planer with a large ferromagnetic component.
Fig.\,\ref{ma}(b) shows the magnetic order at $U=4$ eV and $\alpha=0.2$. 
It is clear from this figure that
 raising the strength of Coulomb interaction at finite $\alpha$
suppresses the ferromagnetic component.
The magnetic order at the same value of $U$ ($4$ eV)
and a higher $\alpha$, $\alpha=1$,
is shown in Fig.\,\ref{ma}(c),
where a stronger SOC has changed
 the orientations and reduced the sizes of the individual moments. 

\section{DISCUSSION AND SUMMARY}
The SOC is an essential ingredient in numerous exciting phenomena including spintronics
and topological insulators. 
However, in transition metal oxides with 3d orbitals such as high temperature cuprates, the SOC has been ignored, while
the strong correlation represented by Hubbard interaction determines their physical properties.
Very recently, iridates with 5d-orbital has been a topic of much attractive research due to intriguing combined effects of
the SOC and Hubbard interaction. It was found that the SOC in iridates is unusually strong, which differs from
other 5d compounds such as Re-oxides \cite{ClancyArxiv12056540}
even though the atomic SOC should be similar for 
Ir and Re.

A set of iridates is the perovskite iridates forming Ruddlesden Popper series from single layer Sr$_2$IrO$_4$ to
three dimensional structure SrIrO$_3$. While both single layer and bilayer iridates exhibit a magnetic insulating behavior,
SrIrO$_3$ shows a metallic phase with a small number of charge carriers. Given that the SOC and Hubbard interaction are local,
their strengths should be similar in this series, and thus
it was suggested that the bandwidth should control metal-insulator transition as the number of layers changes in perovskite iridates.\cite{MoonPRL08}
Here we show that the metallicity
is innate 
 to the lattice structure of three dimensional orthorhombic perovskites in addition
to a large SOC. Due to this combined effect, there are tiny hole and electron Fermi pockets with small density of states, which 
in turn makes Hubbard interaction less efficient in SrIrO$_3$. Due to strong SOC, the magnetic field dependence of physical properties would be interesting to study.

We investigate an overall phase diagram of the orthorhombic perovskite structure (space group Pbnm) in $U$ vs. SOC using density functional theory. The computation is based on SrIrO$_3$, where tuning $U$ and SOC (by changing $\alpha$) allows us to explore other possible phases nearby non-magnetic semimetal in isostructural systems.
Three phases -- non-magnetic metal/semimetal, magnetic metal, and magnetic insulator -- were found 
by tuning $U$ and SOC.
At smaller $\alpha$, a magnetic metal is always found, which is similar to SrRuO$_3$. While Ru$^{4+}$ has 4 electrons
at the outer shell and thus the chemical potential is different from SrRhO$_3$, the bands near the Fermi level are 
strongly mixed leading to a similar phenomena. 
Indeed, earlier electronic calculation on SrRuO$_3$
reported it a ferromagnetic metal.
At $\alpha > 0.3$ and $U < 1.5eV$, the system becomes non-magnetic metal which resembles the ground state of SrRhO$_3$.
Indeed, our computations of the electronic structure of SrRhO$_3$ shows that it is similar to the one found at $\alpha= 0.4$
close to the instability towards magnetic metallic phase.
While Rh and Ru are next to each other in the periodic table, our results imply that the SOC must have a stronger effect on SrRhO$_3$ than
SrRuO$_3$, and agree with an earlier suggestion that SrRhO$_3$ is near a magnetic critical point \cite{YamauraPRB01,ShimuraJSSC92,SinghPRB03}.
Increasing $\alpha$ further, the bands near the Fermi level changes to semimetallic-like, and a stronger $U_c$ is required
for a magnetic insulator.  
The shape of phase boundary between the non-magnetic semimetal and the magnetic insulator is emerged from a line of Dirac node
leading a small density of states near the Fermi level.
A tight binding approach for a series of Sr$_{n+1}$Ir$_{n}$O$_{3n+1}$ has found the same conclusion that $U_c$ is 
larger for $n=\infty$ than $n=1$ or $n=2$.\cite{CarterUnpublished}

In summary, we have studied the interplay between the SOC and Hubbard interaction in orthorhombic perovskite oxide
with the point group symmetry of Pbnm. Three different phases were identified. A magnetic metal with a finite ferromagnetic
component found in smaller SOC at all values of $U$ 
investigated
 in this study.
Increasing the SOC leads to a phase transition to a non-magnetic metal for small $U$, and to a magnetic insulator for large $U$.
The detailed band structures near the Fermi level in these phases strongly depend on the strength of the SOC rather than
$U$, unless the interaction $U$ leads to another magnetic phase.
Our study may be useful in understanding different ground states found among isostructural perovskites including SrRuO$_3$,
SrRhO$_3$, and SrIrO$_3$. It also provides a microscopic mechanism for semimetallic behaviour in SrIrO$_3$ distinct
from its sister compounds, Sr$_2$IrO$_4$ and Sr$_3$Ir$_2$O$_7$.

\acknowledgements{
We thank H. Takagi for useful discussions. This work is supported by NSERC of Canada (HYK).
MAZ thanks IDB and CCT \& COT for financial support.
}


\end{document}